\documentclass[aps,prb,amsmath,amssymb,nofootinbib,twocolumn,
superscriptaddress]{revtex4-1}

\usepackage[T1]{fontenc}
\usepackage{color,array}
\usepackage{slashed}
\usepackage{graphicx}
\usepackage{float}
\usepackage{epsfig}
\usepackage{bm}
\usepackage{psfrag}
\usepackage{hyperref}
\graphicspath{{fig/}}

\newcommand{\dg}{\dagger}
\newcommand{\up}{\uparrow}
\newcommand{\down}{\downarrow}
\newcommand{\ket}[1]{\left| #1\right\rangle}
\newcommand{\bra}[1]{\left\langle #1\right|}

\newcommand{\tp}{t_\perp}
\newcommand{\Vp}{V_\perp}

\newcommand{\llangle}{\langle\!\langle}
\newcommand{\PS}{\tilde{S}}
\newcommand{\rrangle}{\rangle\!\rangle}
\DeclareMathOperator{\Tr}{Tr}

\newcommand{\stack}[2]{\begin{matrix}#1\\#2\end{matrix}}
\newcommand{\uod}{\ket{\stack\up\down}}
\newcommand{\dou}{\ket{\stack\down\up}}
\newcommand{\udoud}{\ket{\stack{\up\down}{\up\down}}}
\newcommand{\udo}{\ket{\stack{\up\down}\cdot}}
\newcommand{\oud}{\ket{\stack\cdot{\up\down}}}

\begin{document}

\title{Nonlocal density interactions in auxiliary-field  quantum Monte Carlo simulations: \\ application to the square lattice bilayer and honeycomb lattice}
\date{\today}

\author{Michael Golor}
\affiliation{\mbox{\it Institut f\"ur Theoretische Festk\"orperphysik, JARA-FIT and JARA-HPC, RWTH Aachen University, 52056 Aachen, Germany}}

\author{Stefan Wessel}
\affiliation{\mbox{\it Institut f\"ur Theoretische Festk\"orperphysik, JARA-FIT and JARA-HPC, RWTH Aachen University, 52056 Aachen, Germany}}

\begin{abstract}
We consider an efficient scheme to simulate fermionic Hubbard models with nonlocal density-density interactions in two dimensions, based on bond-centered auxiliary-field quantum Monte Carlo.  
The simulations are shown to be sign-problem free within a finite, restricted parameter range.
Using this approach, we first study the Hubbard model on the  half-filled square lattice bilayer, including an interlayer repulsion term in addition to the local repulsion, and present the ground state phase diagram within the accessible parameter region.
Starting from the antiferromagnetically ordered state in the absence of interlayer repulsion, the interlayer interactions are found to destabilize the antiferromagnetic order, leading to a band insulator state.
Moreover, for sufficiently strong interlayer tunneling, we also observe the emergence of a direct dimer product state of mixed D-Mott and S-Mott character along the equal coupling line.
We discuss the stability range of this  state within strong-coupling perturbation theory. 
Furthermore, we consider the Hubbard model on the honeycomb lattice with next-nearest-neighbor interactions.
Such an interaction is found to enhance  both charge density  and spin-current correlations within the semimetallic region.
However, inside the accessible parameter region, they do not stabilize long-ranged charge density wave order nor a quantum spin Hall state, and the only insulating state that we observe exhibits long-range antiferromagnetism.

\end{abstract}

\maketitle


\section{Introduction}
\label{sec:intro}
In the field of strongly correlated electrons, the two-dimensional single-band Hubbard model is arguably the most studied model, particularly in the context of the metal-insulator transition and high-$T_c$ superconductivity.  
Even though the long-ranged Coulomb repulsion is considered as screened to a local interaction among electrons, the Hubbard model supports a variety of phases, including (anti-)ferromagnetism and (to a less certain degree) also unconventional superconductivity.  
Adding nonlocal repulsive interactions opens the large chest of extended Hubbard models with even richer phase diagrams.
However, in comparison to the standard Hubbard model, such extended models have been studied far less by numerical means, in particular beyond one spatial dimension. 
Numerical methods that have been used in these instances include exact diagonalization~\cite{ohta1994,kaneko2013}, variational cluster methods~\cite{aichhorn2004}, two-particle self-consistent approaches~\cite{davoudi2006,davoudi2008}, extensions of dynamical mean field theory (DMFT)~\cite{ayral2013,wu2014,vanloon2014} or variational Monte Carlo~\cite{tocchio2014}.

With respect to quantum Monte Carlo (QMC) simulations, such studies have been rather rare~\cite{zhang1989,rademaker2013}.
This is in particular due to the fermionic sign problem, which restricts their applicability to only small finite systems and high temperatures.
In situations, where the sign problem can be eliminated, determinantal QMC methods often provide rather reliable information about emerging phases and the nature of the corresponding phase transitions. 
However, a direct decoupling of extended density-density interactions for each spin-flavor channel, such as employed in Refs.~\onlinecite{zhang1989,rademaker2013}, leads to a sign problem even at half filling on a bipartite lattice, while the sign problem is avoided when only local interactions are considered. 
Recently, in the context of graphene, where interactions on the honeycomb lattice are expected to be truly long-ranged and thus described by a Coulomb model, more suitable QMC methods have been employed~\cite{ulybyshev2013,hohenadler2014}.
There, extended interactions are treated by coupling the electronic density to a (continuous) scalar field, over which a Monte Carlo sampling is performed.
Concerning the absence of the sign problem, this method requires -- in addition to the usual particle-hole symmetry -- the interaction strength to decrease sufficiently rapidly with distance, in a sense that will be specified further below. 

Here, we consider an alternative approach to include extended density-density interactions in determinantal QMC simulations that suffers from similar constraints as the aforementioned method, but which may be more simple to implement for models with short-ranged nonlocal interactions only. 
After introducing the approach in Sec.~\ref{sec:method}, we apply our implementation to extended Hubbard models on two different lattice geometries,  both of which have been discussed  to some extent in the recent literature:

As a first application, we consider in Sec.~\ref{sec:bilayer} the Hubbard model on the square lattice bilayer with interlayer repulsion at half-filling. 
The phase diagram of this model with only local repulsions has been revisited in a previous publication~\cite{golor2014}, with the conclusion that the interacting system is either an antiferromagnetic insulator or a band insulator, primarily decided by the ratio of interlayer to intralayer hopping.
The addition of interlayer interactions was  discussed previously in connection with exciton condensation~\cite{eisenstein2004}, and has been studied with determinantal QMC~\cite{rademaker2013} and exact diagonalization~\cite{kaneko2013} recently, with a focus on the impact on interlayer pairing.
A recent cellular DMFT study~\cite{vanhala2015} considered attractive interactions and found signs of charge-ordered and superfluid phases.

In Sec.~\ref{sec:bilayer}, we investigate the effects of an interlayer repulsion on the phases of the half-filled system, and observe the destruction of antiferromagnetic order within the accessible parameter region. 
We then elaborate on the nature of the band insulator phase and exhibit the existence of an exact dimer product state inside the strong coupling region. 
We rationalise these findings based on large-coupling effective theories~\cite{tsuchiizu2002}.
In a related Hubbard model with both interlayer density interactions and interlayer Heisenberg exchange, a previous QMC study~\cite{capponi2004} revealed the existence of a current-carrying phase.
But also therein, the sign problem confined the explorable parameter space, and the interaction range that we consider here has not be explored in that previous study. 

In Sec.~\ref{sec:honeycomb}, we turn our attention to the Hubbard model on the honeycomb lattice with next-nearest neighbor (NNN) interactions.
Such a model has been proposed in a mean-field study~\cite{raghu2008} as an example for an interaction-driven topological Mott insulator state.
According to that analysis, the extended interactions induce an effective spin-orbit coupling, which then triggers a quantum spin Hall (QSH) phase.
Subsequent studies~\cite{grushin2013,garcia2013,daghofer2014,djuric2014,capponi2015,motruk2015,scherer2015} concentrated on the spinless variant of that model and the existence of a corresponding quantum anomalous Hall phase, with most results pointing against its existence in that model, and instead finding a direct transition from a semimetal regime to a phase with charge density wave (CDW) order. 
Using our QMC approach, we perform large-scale simulations of the less-studied spinful model.
We find that the accessible parameter range prevents us from observing transitions to new phases.
However, we  observe tendencies towards a CDW phase and an increase of QSH-compatible spin currents within the semimetallic phase prior to the instability towards antiferromagnetism. 
Final conclusions are presented in Sec.~\ref{sec:concl}.



\section{Model and Method}
\label{sec:method}
In the following, we are interested in the  simulation of  Hubbard models with additional nonlocal density-density interactions, as described by the Hamiltonian
\begin{align}
  \label{eq:ext-hubb}
   H&=H_t+H_U+H_V,\nonumber\\
H_t&=-\sum_{\langle ij\rangle\sigma}t_{ij}c^\dg_{i\sigma}c^{}_{j\sigma},\nonumber\\   
 H_U&=\frac U2\sum_{i}(n_i-1)^2,\nonumber\\
   H_V&=\frac12\sum_{i\neq j}V_{ij} (n_{i}-1)(n_{j}-1),
\end{align}
where $n_i=n_{i\up}+n_{i\down}$ is the local density operator, $U$ denotes local repulsion and the matrix elements $V_{ij}$ determine the density interaction between lattice sites $i$ and $j$.
Furthermore, $H_t$ denotes the free kinetic part of the Hamiltonian with hopping amplitudes $t_{ij}$. 
Such models have been recently studied with QMC methods in the context of long-ranged Coulomb models on honeycomb lattices\cite{ulybyshev2013,hohenadler2014}, by coupling the electron density to an auxiliary continuous scalar field.
The sign problem is absent for a positive definite interaction matrix $V_{ij}$ (with $V_{ii}=U$).
In this section we employ an alternative approach, which is particularly suited to readily include short-ranged interactions in auxiliary-field QMC implementations that already support simulations of standard Hubbard models.
The procedure is explained  here for the \textit{projective} auxiliary-field QMC method\cite{assaad_det_qmc_2008}, which allows for the calculation of ground state properties and is based on the imaginary-time evolution of a trial wave function $\ket{\Psi_T}$ to the ground state,
\begin{equation}
  \label{eq:qmc-proj}
  \langle O\rangle=\lim_{\Theta\to\infty}\frac{\bra{\Psi_T}e^{-\Theta H/2}\,O\,e^{-\Theta H/2}\ket{\Psi_T}}{\bra{\Psi_T}e^{-\Theta H}\ket{\Psi_T}},
\end{equation}
at which a large class of observables $O$ can then be measured.
The imaginary-time propagation is split up into $M$ elementary increments of size $\Delta\tau=\Theta/M$ via a Trotter-Suzuki decomposition,
\begin{equation}
  \label{eq:suzuki}
  e^{-\Theta H}=\left[e^{-\Delta\tau H_t}e^{-\Delta\tau H_U}e^{-\Delta\tau H_V}\right]^M+{\cal O}(\Delta\tau^2).
\end{equation}
The propagation by the interacting parts $H_U$ and $H_V$ will be rendered feasible via two successive Hubbard-Stratonovich (HS) decouplings,
\begin{equation}
  \label{eq:hs-trafo}
  e^{-\frac12 A^2}=\frac1{2\pi}\int_{-\infty}^\infty\!d\phi\,e^{-\frac12\phi^2}\,e^{i\phi A},
\end{equation} 
at the cost of two new auxiliary \textit{real} bosonic fields $\phi_U(i,\tau)$ and $\phi_V(ij,\tau)$, over which the Monte Carlo sampling is eventually performed.
A HS transformation requires the interaction term to be present in quadratic form $A^2$, which in general can be achieved in different ways.
The particular choice of HS transformations is crucial for the nature of the sign problem\cite{wu2005} and will be explained in the following.

In a first step, the extended density interaction is rewritten in a form that couples the auxiliary field $\phi_V$ to the particle density on the corresponding bond,
\begin{align}
  \label{eq:ext-dec}
(n_i-1)(n_j-1)=&\tfrac12(n_i+n_j-2)^2-\tfrac12(n_{i}-1)^2\nonumber\\
&\qquad-\tfrac12(n_j-1)^2.
\end{align}
After identifying $A^2=\Delta\tau V_{ij}(n_i+n_j-2)^2$ and performing the HS transformation, the propagation by $H_V$ at time slice $m_\tau$ in Eq.~(\ref{eq:suzuki}) will be replaced by
\begin{align}
 e^{-\Delta\tau H_V}\propto\prod_{i,j}\!\int\!d\phi_V&(ij,\tau)\,e^{-\frac12\phi_V^2(ij,\tau)}\nonumber\\
&\times \,e^{i\sqrt{\Delta\tau V_{ij}}(n_i+n_j-2)\phi_V(ij,\tau)}.
\end{align}
Beside the desired quadratic term, other terms appear in Eq.~(\ref{eq:ext-dec}) that contribute to the local Hubbard interaction and hence will be absorbed into $H_U$.
For general $V_{ij}$, this will result in new renormalized values $\tilde U_i$.
However, let us assume in the following, that the extended interaction strength is constant and the interacting bonds form a lattice with connectivity $z_V$. In this case the local interaction will be renormalized to $\tilde U=U-z_VV$.

Since the local interaction term is already of quadratic form, we can straightforwardly perform a second HS transformation and couple the local density to the field $\phi_U(i,\tau)$,
\begin{align}
 e^{-\Delta\tau H_U}\propto\prod_{i}\!\int\!d\phi_U(i,\tau)\,e^{-\frac12\phi_U^2(i,\tau)}\,e^{i\sqrt{\Delta\tau\tilde U}(n_i-1)\phi_U(i,\tau)}
\end{align}
As it turns out, the sign of the renormalized $\tilde U$ determines whether a sign problem occurs or not.
More precisely, sign problem-free calculations require a nonnegative value of $\tilde U$.
The proof is analogous to the one given in  Ref.~\onlinecite{hohenadler2012}: Due to the $SU(2)$ symmetry of Eq.~(\ref{eq:ext-hubb}), the configuration weight can be factored into  spin components, $W[\{\phi_U\},\{\phi_V\}]=\prod_{\sigma=\up,\down}W_\sigma$, with
\begin{widetext}
  \begin{align}
    W_\sigma\big[\{\phi_U\},\{\phi_V\}\big]=\Tr\left[e^{\Theta E_T}\exp\left\{-\Theta\sum_{ij} c_{i\sigma}^\dagger[h_T]_{ij}c_{j\sigma} \right\}\prod^M_{m_\tau=1}\exp\left\{-\Delta\tau(-t)\sum_{\langle ij\rangle} c_{i\sigma}^\dagger c_{j\sigma}+\text{h.c.}\right\}\right.\nonumber\\
\hfill\left.\times\exp\left\{i\sqrt{\Delta\tau\tilde U}\sum_i(n_{i\sigma}-\tfrac12)\phi_U(i,m_\tau\Delta\tau)\right\}\exp\left\{i\sqrt{\Delta\tau V_{ij}}\sum_{ij}(n_{i\sigma}+n_{j\sigma}-1)\phi_V(ij,m_\tau\Delta\tau)\right\}\right],
  \end{align}
\end{widetext}
where the Hamiltonian $H_T=\sum_{ij\sigma} c_{i\sigma}^\dagger[h_T]_{ij}c_{j\sigma}$ is used for the generation of the trial wave function, with $H_T\ket{\Psi_T}=E_T\ket{\Psi_T}$. 
The sign problem is absent, if one can find a canonical transformation that connects both weights as $W_\down=W^*_\up$, ensuring a positive total weight $W=W_\up W^*_\up=|W_\up|^2>0$.
The $SU(2)$ symmetry together with a particle-hole symmetry allows for a canonical transformation $c_{i\up}\to(-1)^i c_{i\down}^\dagger$, where the factor $(-1)$ only contributes on one of the two sublattices.
Its application to $W_\up$ and a subsequent complex conjugation will -- except for the spin flip -- not affect the trial wave Hamiltonian and the kinetic contribution, as long as both describe bipartite systems.
The particle density will transform as $n_{i\up}\to1-n_{i\down}$, introducing a minus sign in the corresponding exponentials. This minus sign will only get canceled by the complex conjugation, if both terms are purely imaginary, which is indeed the case for
\begin{equation}
  \label{eq:sign-condition}
  \tilde U=U-z_VV\ge 0.
\end{equation}
In the case of nearest-neighbor (NN) repulsion only, it turns out that this constraint is identical to the positive (semi)definiteness of the interaction matrix $V_{ij}$, which is required for the approach employed in Refs.~\onlinecite{ulybyshev2013,hohenadler2014}.
This is no coincidence, and in fact, our approach can be obtained within this framework.
While in the approach of  Refs.~\onlinecite{ulybyshev2013,hohenadler2014}, the scalar field couples to the largest modes of the global interaction matrix dynamically during the simulation, we  carried out the diagonalization locally, and thereby couple  to these largest eigenvectors in terms of  local, bond-centered auxiliary fields.
For further extended interactions, such as the case of NNN repulsion on the honeycomb lattice considered in Sec.~\ref{sec:honeycomb},  we can in fact arrive at a less restricted interaction range than implied by Eq.~\ref{eq:sign-condition}, by employing an appropriate local decoupling of the interaction terms, as discussed in detail in Sec.~\ref{sec:honeycomb}.
Finally, we note that in practice, we employ a discrete version of the HS auxiliary field decoupling instead of the 
continuous field considered in the above derivation, e.g., the $SU(2)$ symmetric HS decoupling of 
Ref.~\onlinecite{hohenadler2012}. 



\section{Square Lattice Bilayer with interlayer repulsion}
\label{sec:bilayer}
In this section, we examine the half-filled square lattice bilayer with finite interlayer repulsion, as described by the Hamiltonian
\begin{align}
  \label{eq:sqb-ham}
    H=-t\sum_{\langle ij\rangle\sigma\lambda} c^\dg_{i\lambda\sigma}c^{}_{j\lambda\sigma} -\tp\sum_{i\sigma}(c^\dg_{i1\sigma}c^{}_{i2\sigma}+c^\dg_{i2\sigma}c^{}_{i1\sigma})\nonumber\\
+\frac U2\sum_{i\lambda}(n_{i\lambda}-1)^2+\Vp\sum_i (n_{i1}-1)(n_{i2}-1),
\end{align}
where $\lambda\in\{1,2\}$ denotes the layer index, $\tp$ the interlayer hopping amplitude and $\Vp$ the interlayer density-density repulsion.
The interaction connectivity of the nonlocal interaction for this model is $z_{\Vp}=1$, hence from the condition in Eq.~(\ref{eq:sign-condition}) it follows that we can simulate this model in the region $\Vp\le U$  without a sign problem.
In contrast, in a recent QMC study of the same in model in Ref.~\onlinecite{rademaker2013}, where a separate, direct HS decoupling of the interlayer repulsion for each spin-spin sector was employed, a sign problem arises for any finite values of $\Vp>0$ already for the half-filled system. 

\subsection{Phase diagram}
\label{sec:sqb-phase-diagram}
\begin{figure}[t]
  \centering
  \includegraphics[width=0.95\columnwidth]{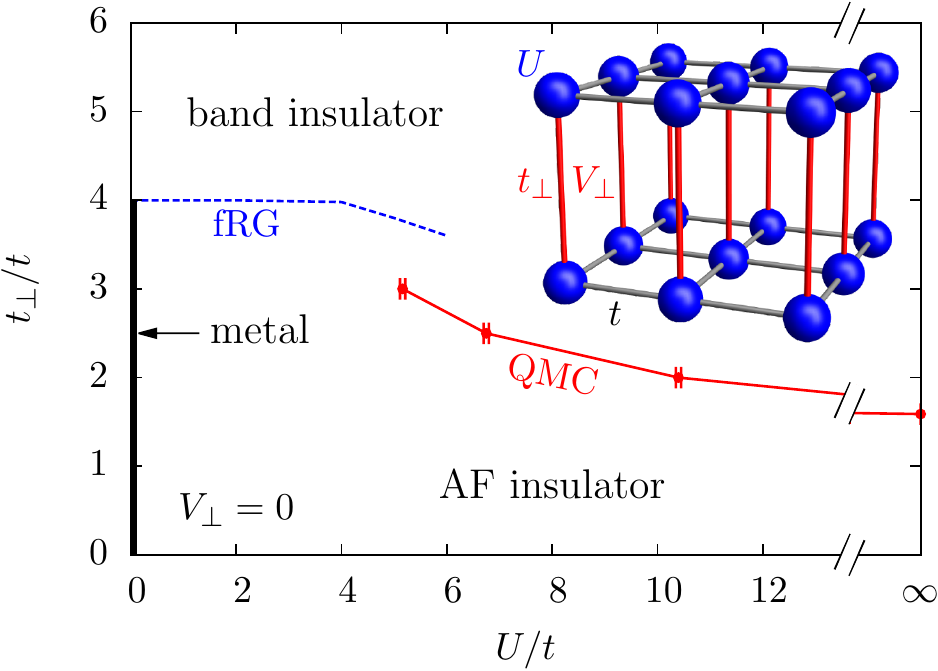}
  \caption{(Color online) Phase diagram of the half-filled Hubbard bilayer in the absence of interlayer interactions ($\Vp=0$), with estimates of the phase boundary between the AF insulator and the band insulator as obtained from QMC and fRG calculations. The inset illustrates the square lattice bilayer geometry. Figure reproduced from Ref.~\onlinecite{golor2014}.}
  \label{fig:sqb-phasediagram-v0}
\end{figure}

In the limiting case of vanishing interlayer interactions, $\Vp=0$, there have been suggestions~\cite{kancharla2007,bouadim2008} for a paramagnetic metallic phase persisting at small values of $U/t$ in the region of low $\tp/t<4$.
However, more recent analyses~\cite{rueger2014,golor2014} argue for a direct onset of an antiferromagnetic (AF) insulator at any finite interaction strength in the low $\tp/t$ region. 
Increasing the interlayer hopping $\tp/t$ however favors singlet formation on the interlayer bonds and eventually drives the system into a band insulator phase.
The $\Vp=0$ phase diagram as obtained from a combined QMC and functional renormalization group (fRG) analysis~\cite{golor2014} is shown in Fig.~\ref{fig:sqb-phasediagram-v0} for reference, while in the following we will study the effects of finite $\Vp>0$ on the phase diagram.

For the  calculations at finite $\Vp$, we employed the projective QMC method along with the extensions introduced in Sec.~\ref{sec:method}, taking the ground state of the non-interacting model as the trial wave function.
We used a projection length $\Theta=60/t$ and a discretization step $\Delta\tau=0.1/t$, chosen sufficiently low such that corresponding systematic errors are below the statistical uncertainties.
The available computational resources allowed us to simulate systems with a maximum linear length of $L=24$, corresponding to $N=2L^2=1152$ lattice sites.
We restrict our analysis on the intermediate-to-strong coupling regime in the following, since QMC studies of the weak-coupling region ($U\lesssim4t$) suffer not only from exponentially small AF order parameters in that region, but also from large finite-size effects due to the Fermi surface structure of the noninteracting tight-binding model on the bilayer square lattice\cite{golor2014}.

Besides the AF and band insulator phases, a charge density wave (CDW) state with alternating electron densities on the two sublattices may be expected to stabilize for sufficiently high values $\Vp/U$.
The relevant  observables for the ordererd states are thus the  structure factors for antiferromagnetic and staggered CDW order, respectively, given by 
\begin{align}
  S_{\rm AF}&=\frac1N \sum_{i,j,\lambda} e^{i\bm Q(\bm r_i-\bm r_j)} \left[\langle\bm S_{i\lambda} \cdot \bm S_{j\lambda}\rangle-\langle\bm S_{i\lambda} \cdot \bm S_{j\bar\lambda}\rangle\right],\nonumber\\
  S_{\rm CDW}&=\frac1N \sum_{i,j,\lambda} e^{i\bm Q(\bm r_i-\bm r_j)} \left[\llangle n_{i\lambda} n_{j\lambda}\rrangle-\llangle n_{i\lambda} n_{j\bar\lambda}\rrangle\right],
\end{align}
where $\bm S_{i\lambda}$ is the spin vector operator, $\bm Q=(\pi,\pi)$ is the ordering vector and $\bar\lambda$ denotes the opposite layer of $\lambda$, i.e., $\bar1=2$, $\bar2=1$. 
The cumulant is defined as $\llangle OP\rrangle=\langle OP\rangle-\langle O\rangle\langle P\rangle$ for two arbitrary operators $O,P$.

\begin{figure}[t]
  \centering
  \includegraphics[width=\columnwidth]{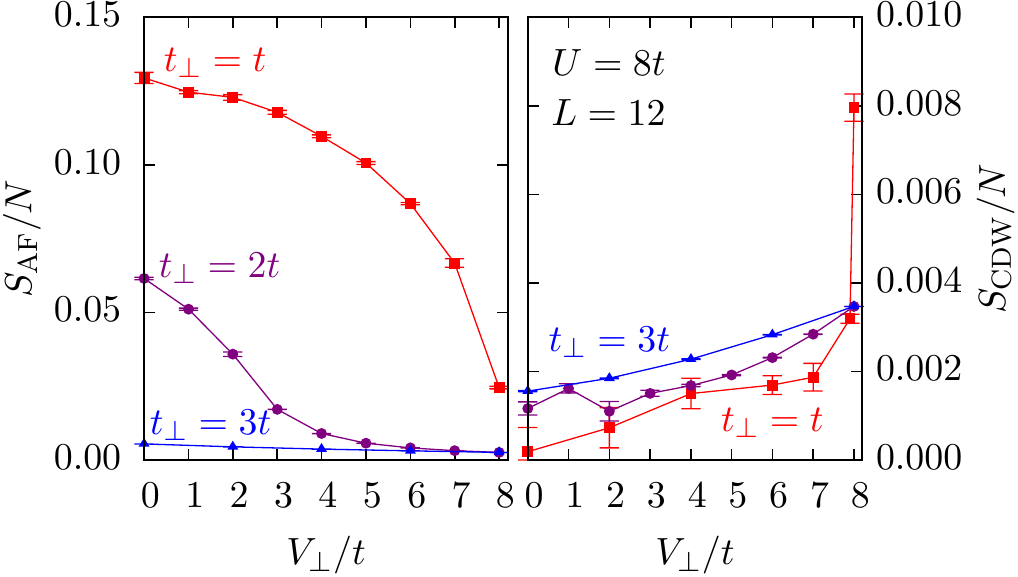}
  \caption{(Color online) Effect of interlayer repulsion on the AF (left) and CDW (right) structure factors for different interlayer hoppings $\tp/t$. The data shown is obtained from simulations of a $L=12$ system at $U=8t$.}
  \label{fig:sqb-sf}
\end{figure}

\begin{figure}[t]
  \centering
  \includegraphics[width=0.95\columnwidth]{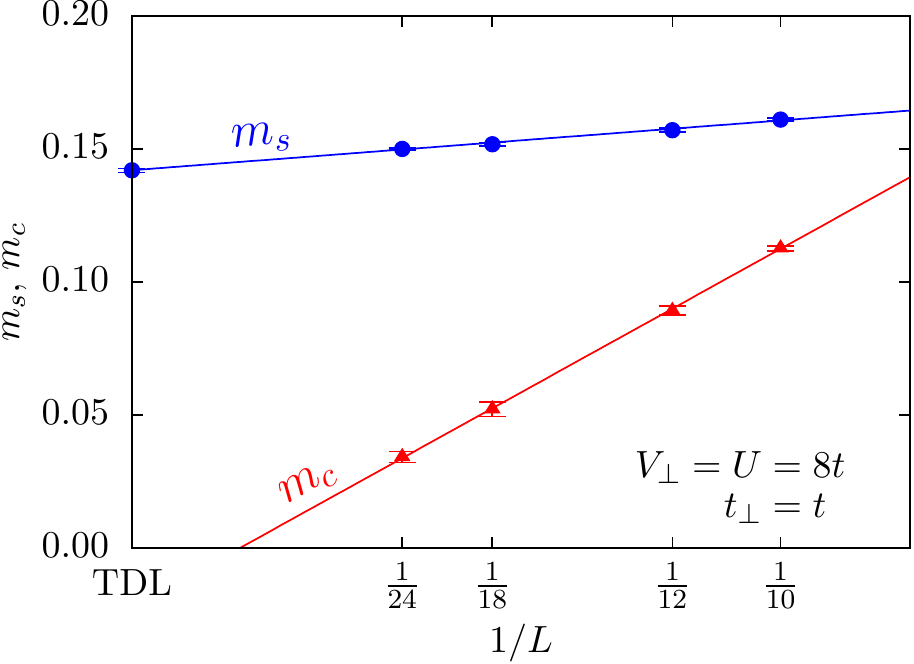}
  \caption{(Color online) Finite-size behavior of spin (disks) and charge (triangles) order parameters along with linear extrapolations in $1/L$ to the thermodynamic limit after the observed jump of the CDW structure factor at $\Vp=U=8t$, $\tp=t$.}
  \label{fig:sqb-fss}
\end{figure}

\begin{figure}[t]
  \centering
  \includegraphics[width=0.95\columnwidth]{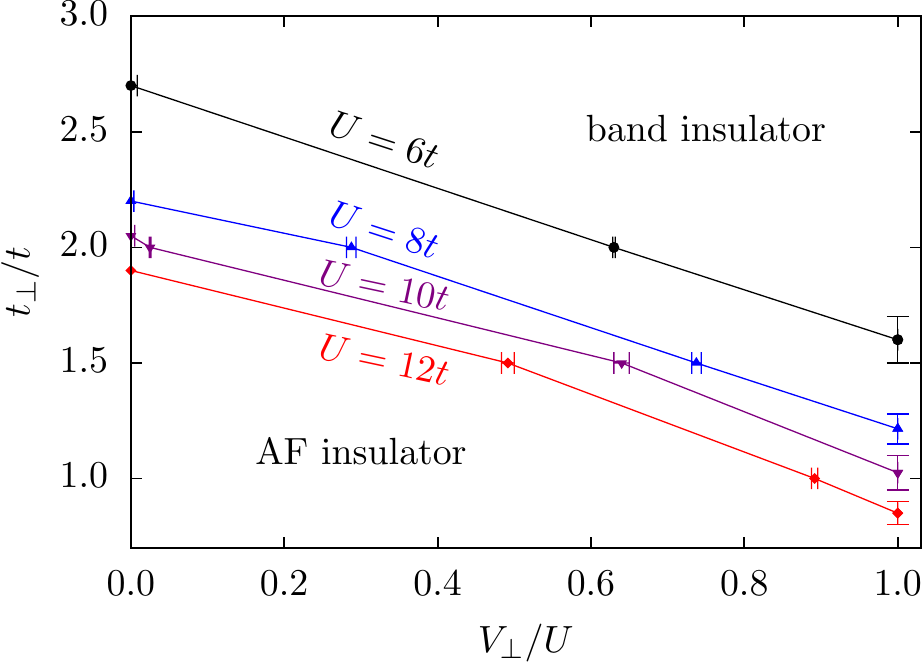}
  \caption{(Color online) Phase boundaries between the AF and the band insulator phase for different values of the  local interaction strength (lines are guides to the eye). For $\Vp<U$, the transition is described by the three-dimensional Heisenberg $O(3)$ universality class. At $\Vp=U$, the onset of magnetization is estimated by finite-size extrapolations.}
  \label{fig:sqb-phasediagram-tv}
\end{figure}

In Fig.~\ref{fig:sqb-sf}, both these structure factors are shown as a function of the interlayer repulsion for different interlayer hopping strenghts, as obtained for a system with $L=12$, $U=8t$.
Increasing the interlayer repulsion can be seen to strongly suppress the AF tendencies, while it only weakly augments the CDW signal.
For $\tp=t$, a doubling of the CDW structure factor from $\Vp=7.9t$ to $\Vp=8t$ can be observed, hinting at a possible first-order transition at this point.
However, a careful finite-size extrapolation to the thermodynamic limit of the corresponding order parameters $m_s=\sqrt{S_{\rm AF}/N}$ and $m_c=\sqrt{S_{\rm CDW}/N}$ supports a finite AF order parameter (Fig.~\ref{fig:sqb-fss}), and thus the system remains in the AF phase also at this point. 
In the opposite case of large interlayer hopping $\tp=3t$, the system resides within the band insulator phase already at zero interlayer interaction (cf.~also Fig.~\ref{fig:sqb-phasediagram-v0}), which is found to be stable for all accessible values of $\Vp$.
Only for the intermediate case of $\tp=2t$, do we observe a phase transition between the AF and band insulator induced by $\Vp$.
Performing a finite-size scaling analysis of the AF structure factor, with the scaling ansatz
\begin{align}
  S_{AF}/N=L^{-\frac{2\beta}\nu}\,{\cal F}_S(g L^{\frac1\nu}),\quad g=\frac{\Vp-V_{\perp,c}}{V_{\perp,c}},
\end{align}
and using the critical exponents of the three-dimensional Heisenberg $O(3)$ universality class~\cite{campostrini2002}, one can indeed locate the transition point at $V_{\perp,c}=2.24(3)t$.
\begin{figure}[t]
  \centering
  \includegraphics[width=0.95\columnwidth]{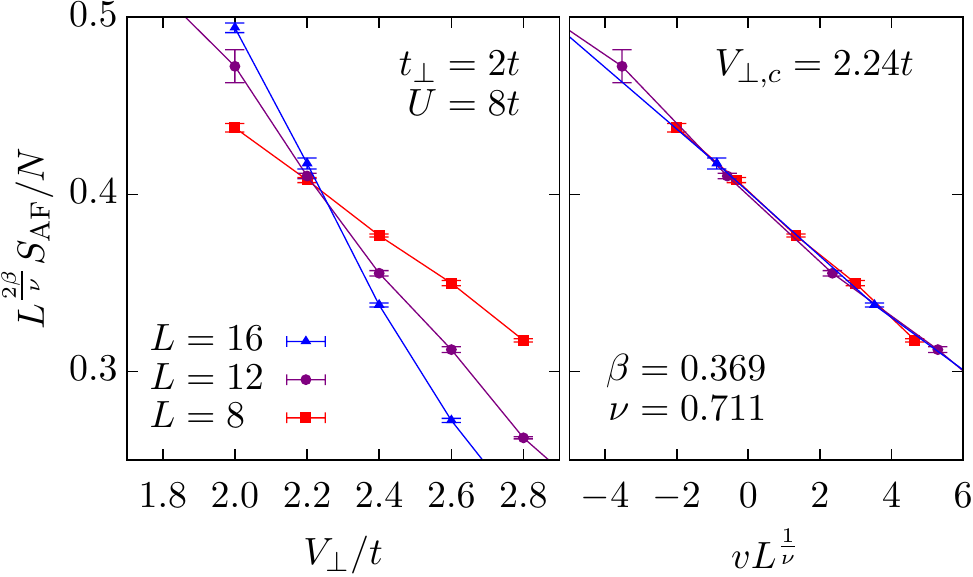}
  \caption{(Color online) Finite-size scaling analysis of the phase transition at $U=8t$, $\tp=2t$. Left panel: Crossing plot for determining the critical $\Vp$. Right panel: Data collapse from using the critical exponents of the three-dimensional Heisenberg $O(3)$ universality class. }
  \label{fig:sqb-coll}
\end{figure}
The finite-size data exhibit a crossing point at this critical value of $\Vp$, and furthermore produce a good data collapse in the critical region, as shown in the left and right panel of Fig.~\ref{fig:sqb-coll}, respectively. 
Repeating the above scaling analysis for other values of the local interaction $U$ results in the phase diagram shown in Fig.~\ref{fig:sqb-phasediagram-tv}.

Within the accessible region of $\Vp\le U$, we observe only an AF phase and a band insulator phase with a phase boundary that moves towards smaller interlayer hopping strength upon increasing $U/t$.
This qualitative behavior presumably extends also into the weak coupling regime.

\subsection{Direct dimer product state for large $\Vp= U$ }
\label{sec:comp-with-effect}
While analysing the equal coupling case, $\Vp=U$, we observed a rather peculiar behavior in the QMC dynamics. 
We find that above a critical value of the interlayer hopping $t_{\perp,c}$ and after only a short thermalization period, all observables appear to be fixed to certain values, and exhibit no more fluctuation in the further course of the simulation.
Depending on the type of observable, the actual values are  even independent of the choice of $U/t$ and $\tp/t$.
For example, the interlayer spin-spin and density-density correlations  take on values $\bm S_{i1}\cdot\bm S_{i2}=-0.375$ and $n_{i1}n_{i2}=0.5$, respectively.
Most remarkably, all correlations between sites that do not share an interlayer bond vanish.
This observation strongly suggests that in these cases the ground state forms a direct dimer product state (DDPS), located on the interlayer bonds.

\begin{figure}[t]
  \centering
  \includegraphics[width=0.95\columnwidth]{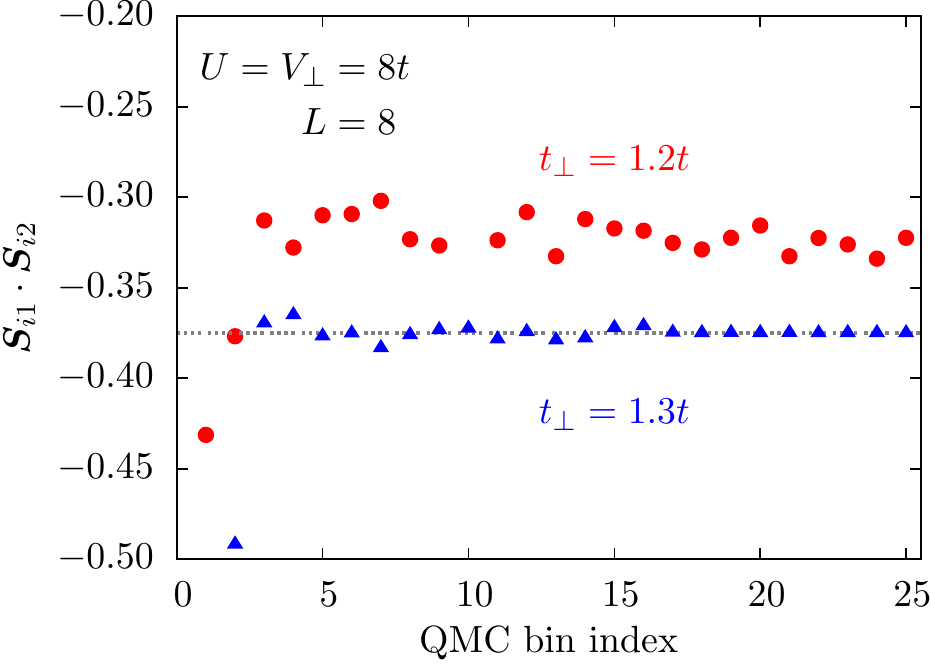}
  \caption{(Color online) QMC dynamics of the interlayer spin-spin correlations at equal coupling $U=\Vp=8t$ for $\tp=1.2t,1.3t$. The dashed horizontal line is set at a value of the ordinate of -0.375. For values $\tp\gtrsim1.3t$, the ground state is given by the DDPS defined in Eq.~(\ref{eq:prod-state}).}
  \label{fig:mc-obs}
\end{figure}

The precise nature of the DDPS can indeed be accessed by an effective theory in the strong-coupling regime.
Such a description has been employed in Ref.~\onlinecite{tsuchiizu2002} for the case of the extended Hubbard model on the two-leg ladder, and can be readily generalized to the two dimensional case.
In the following, the derivation will only be sketched and for a more detailed discussion we refer to Ref.~\onlinecite{tsuchiizu2002}.

In the limit of vanishing hoppings $t,\tp\to0$, the wave function can be written as a product of dimer states residing on the interlayer bonds.
In particular for comparable coupling strengths, $\Vp\approx U$, one then finds that only two such states contribute to the ground state manifold, which are the so-called \textit{D-Mott} and \textit{S-Mott} states,
\begin{align}
  \ket D_i = \frac1{\sqrt 2}\left[\uod_i - \dou_i\right]\!,\;& \ket S_i = \frac1{\sqrt 2}\left[\udo_i + \oud_i\right]\!,\nonumber\\
\text{with}\quad \udoud_i\!=&\,c^\dagger_{i1\up} c^\dagger_{i1\down} c^\dagger_{i2\up} c^\dagger_{i2\down}\ket0.
\end{align}
While the  D-Mott state describes a spin singlet, the S-Mott is a symmetric state with strong  charge fluctuation.
These states are now identified as the eigenstates of a pseudospin operator $\PS^x_i=n_{i1}n_{i2}-\frac12$, with $\PS^x_i\ket D_i=\frac12\ket D_i$ and $\PS^x_i\ket S_i=-\frac12\ket S_i$.
Treating finite hopping terms in second-order perturbation theory and subsequently performing a sublattice spin rotation, $\PS^{x,y}_i\to(-1)^i\PS_i^{x,y}$, yields an effective two-dimensional XXZ spin model with a staggered transverse field $h_x$ and a uniform longitudinal field $h_z$,
\begin{align}
\label{eq:eff-spin-model}
  H=J\sum_{\langle ij\rangle}(\PS^x_i\PS^x_j+\PS^y_i\PS^y_j+\Delta\PS^z_i\PS^z_j)\nonumber\\
-\sum_i((-1)^ih_x\PS^x_i+h_z\PS^z),
\end{align}
with effective parameters $J=\frac{3t^2}U$, $\Delta=\frac53$, $h_x=U-\Vp$ and $h_z=4\tp$.

Here, we are interested in the regime of equal coupling, $\Vp=U$, and strong interlayer hopping $\tp$, which translates in the effective XXZ model to a vanishing transversal field $h_x=0$ and a large longitudinal field $h_z$, respectively.
In this limit, the XXZ spin model has been studied thoroughly~\cite{kohno1997,hebert2001,yunoki2002} -- in some of these references in an equivalent formulation as a hard-core boson model -- and the phase diagram is well understood.
For zero or weak longitudinal field $h_z$, the model is in an AF ground state, with the staggered magnetization aligned in the  $z$-direction due to the easy-axis anisotropy.
Increasing the field strength eventually  triggers a first-order spin-flop transition, which then saturates into a fully polarized state at $h_z=\frac{16}3 J$, corresponding to $\tp=4\frac{t^2}U$ in the original bilayer Hubbard model. 

In terms of the original Hubbard model, the pseudospin state fully polarized in the $\PS^z$-direction corresponds to a product of the local eigenvectors $|\PS^z_i=+\frac12\rangle=(\ket D_i+\ket S_i)/\sqrt2$, i.e., the bilayer Hubbard model ground state is given by the DDPS
\begin{align}
\label{eq:prod-state}
  \ket\Psi = \bigotimes_{i}\frac12\left[\uod_i - \dou_i + \udo_i + \oud_i\right].
\end{align}
Indeed, such a state is consistent with the observed values of $\bm S_{i1}\cdot\bm S_{i2}$ and $n_{i1}n_{i2}$, and its energy per site of $E/N=(\Vp-2\tp)/2$ is reproduced by the simulations as well.

\begin{figure}[t]
  \centering
  \includegraphics[width=0.95\columnwidth]{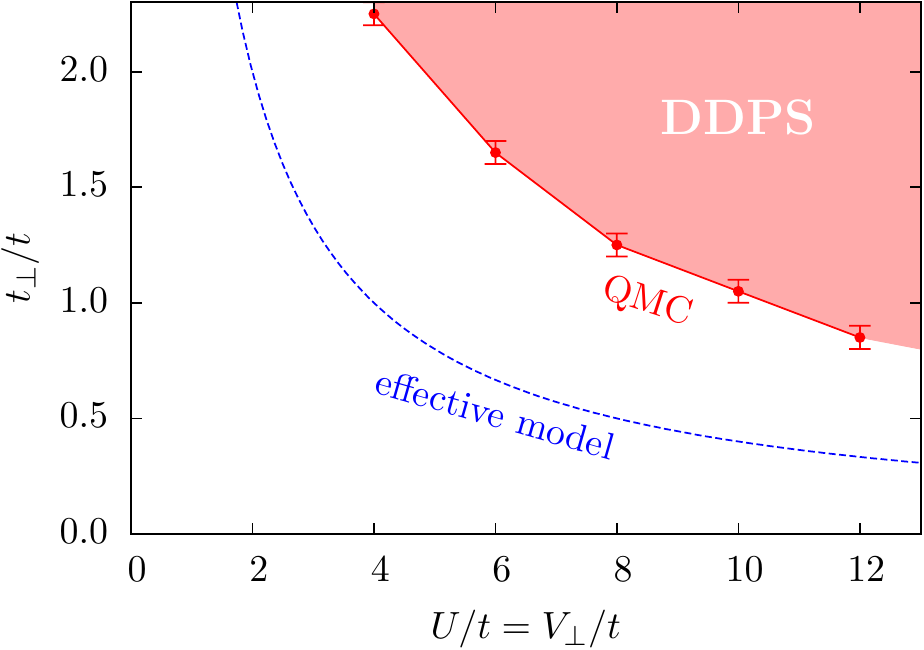}
  \caption{(Color online) Stability line for the direct dimer product ground state of Eq.~(\ref{eq:prod-state}). The red circles represent the QMC results, while the dashed line shows the prediction for the stability line obtained from the effective spin model in Eq.~(\ref{eq:eff-spin-model}).}
  \label{fig:sqb-phasediagram-ps}  
\end{figure}

We can now exploit the particular behavior of this state in the QMC simulation and scan $\tp/t$ in steps of $\Delta\tp=0.1t$ for various values of $\Vp=U$ to determine the critical hopping strength $t_{\perp,c}$ throughout the band insulating phase.
These calculations were performed for linear size $L=8$ and occasionally compared with $L=16,24$ results, where no 
noticeable finite-size effects on the critical hopping were observed.
The resulting stability line is shown in Fig.~\ref{fig:sqb-phasediagram-ps} and separates the DDPS from ground states with finite entanglement between the interlayer dimers.
Within our accuracy, the stability line of the DDPS  falls within the estimated transition region out of the AF 
phase (cf. Fig.~\ref{fig:sqb-phasediagram-tv}).
However, we are not able to discern, whether indeed a direct transition  from the AF phase to the DDPS takes place, or whether there exists a small, intermediate band-insulator region that separates the DDPS from the AF phase.
We noticed however, that  data taken at  $\Vp=U$ for values of $\tp$ close to the transition region within the AF phase does not lead to a data collapse such as those discussed in the previous section.
This may be considered as an indication, that the transition out of the AF phase at $\Vp=U$ is of a different character than the generic transition out of the AF phase into the band insulating regime.
However, within our simulations, we also did not detect clear indications for a (first-order) direct transition from the AF phase to the exact DDPS, which one may expect from the fact that within the DDPS the correlations vanish exactly for all sites not belonging to the same interlayer dimer.
The actual behavior within the transition region to the DDPS at $\Vp=U$ thus remains an open question for possible further research. 

Fig.~\ref{fig:sqb-phasediagram-ps} also contains the phase boundary obtained within the effective XXZ theory, which 
predicts the DDPS to disappear at smaller values of $\tp/t$. 
One needs to note however, that the AF state is not contained in the effective theory's state manifold and thus we do not expect it to describe the transition correctly. 

As a general side note, we remark that it can be shown that the DDPS is an exact eigenstate of the Hubbard model for \textit{any} lattice geometry constructed from dimers, as long as the intradimer repulsion strength coincides with the local repulsion strength.
However, we are not aware of an obvious criterion that determines whether this eigenstate will also be the ground state.

\subsection{Strong interlayer repulsion}
Tuning the interlayer repulsion $\Vp$ beyond the value of the onsite interaction $U$ corresponds in the effective XXZ model in Eq.~(\ref{eq:eff-spin-model}) to a finite value of the transverse magnetic field $h_x=U-\Vp$.
Starting at $h_x=0$ from the direct product state in Eq.~(\ref{eq:prod-state}), which represents an equal-weight superposition of the D-Mott and S-Mott states, a finite transverse field thus enhances the D-Mott (S-Mott) character of the ground state for $h_x>0$, corresponding to $U>\Vp$ ($h_x<0$, corresponding to $\Vp>U$).
For the case of dominant local interactions $U$, this observation is in accord with  results from previous QMC studies of the $U$-only model~\cite{bouadim2008,golor2014}, which at large $\tp/t$ identified a band insulator phase that adiabatically connects to the dimer-singlet state in the large $U$-limit, which is a state of D-Mott character. 

While we cannot perform QMC simulations in the large-$\Vp$ region, we may nevertheless employ a perturbative calculation to access this regime. 
In particular for the region $\Vp\gg U \gg t, \tp$, after a decoupling of the interaction terms in the interlayer rung  states the appropriate basis for the single rung states is set by the two state comprising the S-Mott basis state~\cite{tsuchiizu2002}.
These can be expressed upon introducing a pseudospin notation as follows:
\begin{equation}
\ket{+}_i= \udo_i,\quad \ket{-}_i = \oud_i,
\end{equation}
with corresponding pseudospin operators $\tau^z_i$ and $\tau^x_i$, that act as follows on the above states:
\begin{equation}
\tau^z_i \ket{\pm}_i = \pm \ket{\pm}_i, \quad \tau^x_i \ket{\pm}_i = \ket{\mp}_i. 
\end{equation}
By a calculation similar to the case of the extended Hubbard model on the two-leg ladder~\cite{tsuchiizu2002}, one then obtains within second-order perturbation theory in $t$, $\tp$, an effective quantum Ising model on the two-dimensional square lattice,
\begin{equation}
H=J \sum_{\langle i,j \rangle} \tau^z_i \tau^z_j - h \sum_i \tau^x_i,
\end{equation}
with the effective model parameters given as
\begin{equation}
J=\frac{2t^2}{2\Vp-U}, \quad h=\frac{2\tp^2}{\Vp-U}>0.
\end{equation}
This quantum Ising model exhibits a quantum phase transition for the ratio of $r=h/J$ equal to a critical value of $r_c=3.044(1)$\cite{bloete2002}, between a low-$r$ AF phase and the large-$r$ paramagnetic phase.
In terms of the original Hubbard model, these phases correspond to the CDW  and S-Mott phase, respectively.
Within the perturbative calculation, the stability line of the CDW phase is obtained as
\begin{equation}
\tp/t=\sqrt{r_c}\sqrt{\frac{\Vp-U}{2\Vp-U}},
\end{equation}
so that the CDW state is stable below $\tp/t=\sqrt{r_c/2}\approx1.23$ in the limit $\Vp\gg U$.
For $\tp/t$ beyond this value, the anticipated CDW state with long-range charge order is  replaced by a dimerized state of S-Mott character. 
We expect this state to adiabatically connect to the S-Mott state that emerges for $\Vp\gtrapprox U$, and which we considered above.
We thus find that 
for sufficiently large interlayer hopping  $\tp/t$, 
both the AF and the CDW state are unstable towards dimerization into insulating states with D-Mott and S-Mott character, respectively.
These ground states become degenerate and form the DDPS at equal coupling, $\Vp=U$, 
for sufficiently large values of $\tp/t$, 
as confirmed by our QMC simulations 
in the previous section.



\section{Honeycomb lattice with $V_2$ interactions}
\label{sec:honeycomb}

Here, we consider the half-filled Hubbard model on a honeycomb lattice with second-nearest neighbor density interactions,
\begin{align}
\label{eq:honey-ham}
     H=-t\sum_{\langle ij\rangle\sigma}c^\dg_{i\sigma}c^{}_{j\sigma}&+\frac U2\sum_{i}(n_{i}-1)^2\nonumber\\
   &+\frac{V_2}2\sum_{\llangle ij\rrangle}( n_{i}-1)(n_{j}-1),
\end{align}
where $\llangle ij\rrangle$ denotes next-nearest neighbors (NNN).

\begin{figure}[h]
   \centering
   \includegraphics[width=0.95\columnwidth]{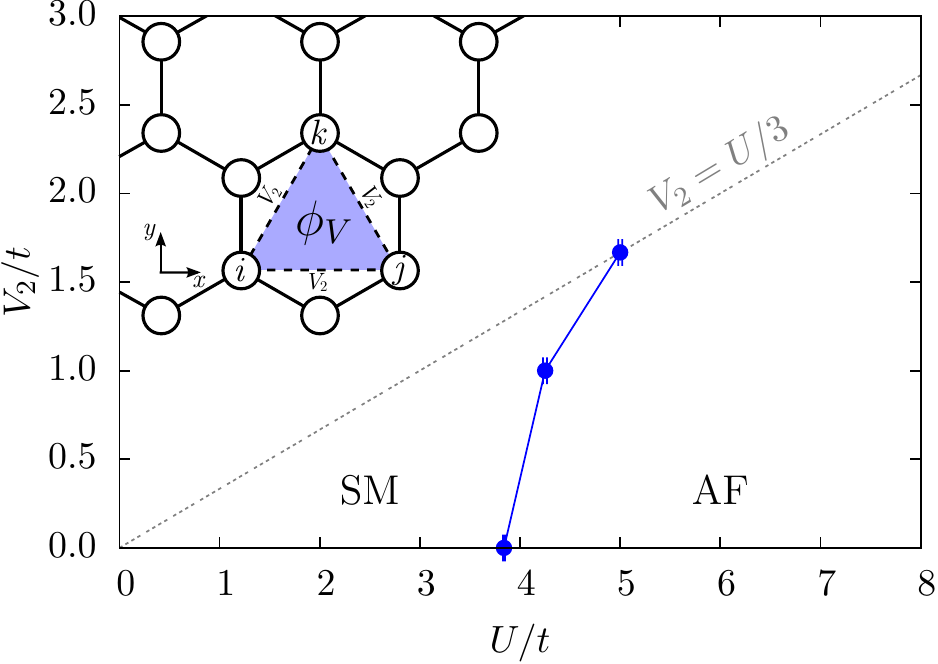}
   \caption{(Color online) Phase diagram of the half-filled Hubbard model on the honeycomb lattice with 
next-nearest-neighbor interactions. The circles indicate the position of the phase boundary as obtained from QMC, while the solid line is a guide-to-the-eye. The gray dashed line marks the boundary of the QMC-accessible region. The inset 
sketches one three-site triangle to whose density the auxiliary field is coupled.}
   \label{fig:honey-phase}
 \end{figure}

Since the coordination number of the $V_2$ interactions is $z_{V_2}=6$, a direct application of the method described in Sec.~\ref{sec:method} would enable us to access a rather small parameter region $V_2\le U/6$ only. However, the particular symmetry of the NNN interactions on the honeycomb lattice allows for a partition into $N$ interacting triangles, each spanned by three lattice sites from one sublattice (cf.~the inset in Fig.~\ref{fig:honey-phase}). This enables a rewriting of the NNN-interaction terms of the involved lattice sites,
\begin{align}
  \label{eq:tri-inter}
(n_i-1)(n_j-1)+&(n_j-1)(n_k-1)+(n_k-1)(n_i-1)\nonumber\\
=&\tfrac12(n_i+n_j+n_k-3)^2-\tfrac12(n_{i}-1)^2\nonumber\\
&-\tfrac12(n_{j}-1)^2-\tfrac12(n_{k}-1)^2,
\end{align}
so that a subsequent HS transformation couples the auxiliary field $\phi_V$ to the total density of each triangle.
This type of coupling proves very beneficial as it doubles the sign problem-free simulation range to $V_2\le U/3$ and moreover reduces the number of auxiliary-field locations from $3N$ to $N$.
Note that the eased parameter restriction does not permit calculations beyond the applicability of the methods used in Refs.~\onlinecite{ulybyshev2013,hohenadler2014}, as it is identical with requiring positive (semi-)definiteness of the interaction matrix.

We implemented the above idea into the projective QMC method and used a projection length of $\Theta=60/t$ and a imaginary-time discretization of $\Delta\tau=0.1/t$. The trial wave function is generated from the noninteracting system's ground state.

The results of our investigation are summarized in the phase diagram in Fig.~\ref{fig:honey-phase}.
At $V_2=0$, the system undergoes a quantum phase transition from a semimetal (SM) to an antiferromagnet (AF) at $U_c=3.843(8)t$\cite{otsuka2015}, described by the chiral-Heisenberg universality class of the Gross-Neveu-Yukawa theory~\cite{herbut2009}.
There has been some debate over a possible intermediate quantum spin liquid phase\cite{meng2010}, however newer results\cite{sorella2012,assaad2013,toldin2015,otsuka2015} obtained on larger clusters are more consistent with a direct transition.

For all accessible finite values $V_2\le U/3$ neither a QSH nor a CDW phase can be detected. Thus, the SM and AF phases remain stable, with only their phase boundary being shifted to larger values ($U_c=5.00(2)t$ on the $V_2=U/3$ line). In the following, we report the details of our calculations for which this phase diagram is obtained.

\subsection{Nature of the large-$V_2$ phase}
\label{sec:honey-phase}

\begin{figure}[t]
  \centering
 \includegraphics[width=0.95\columnwidth]{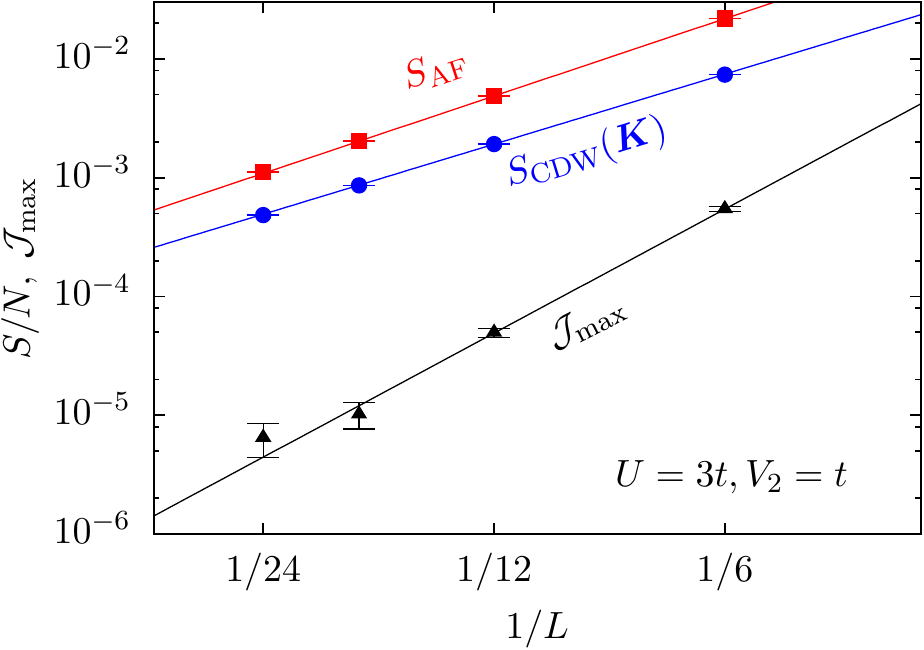}
  \caption{(Color online) Finite-size scaling of AF (red squares) and CDW (blue circles) structure factors and the maximum distance spin current correlation (black triangles) at $U=3t$, $V_2=t$ in a double logarithmic view. An algebraic function $f(1/L)=a L^{-b}$ can be fitted well to the data, implying the persistence of a semimetallic phase at this point.}
  \label{fig:honey-sf-fss}
\end{figure}

Recently, Raghu et al.\cite{raghu2008} suggested that sufficiently strong $V_2$-interactions in this model induce a phase transition from a semimetallic phase to an interacting QSH topological Mott insulator phase.
If confirmed, this system would represent a simple instance of an interaction-induced topological Mott  insulator.
Since then, the spinless counterpart has been studied extensively\cite{grushin2013,garcia2013,daghofer2014,djuric2014,capponi2015,motruk2015,scherer2015}, with most results arguing against the existence of a stable topological phase.
Instead, at large $V_2$ the system apparently enters a \textit{charge-modulated phase}, which can be understood as a frustrated CDW with an ordering vector at the Dirac point $\bm K$.
In the following, we will look in the spinful model for both, charge order and topological order.
\begin{figure}[t]
  \centering
  \includegraphics[width=0.95\columnwidth]{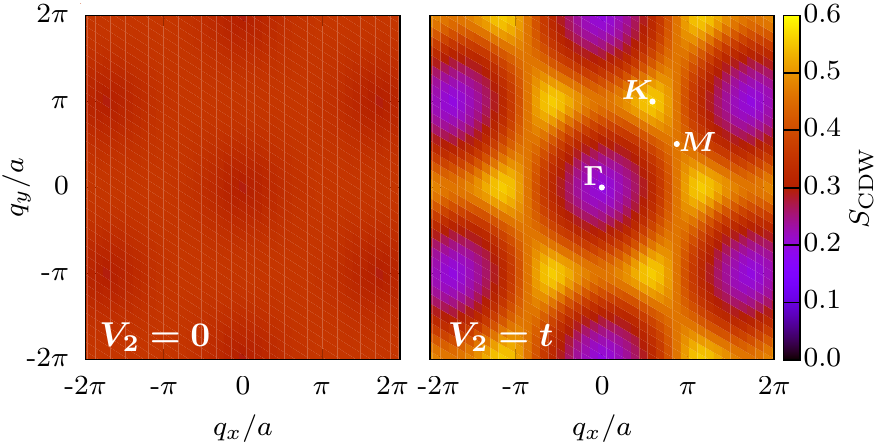}
  \caption{(Color online) Diagonal charge structure factor $S(\bm q)$  at $U=3t$,  $L=18$. The peaks emerge at the $\bm K$-points of the Brillouin zone. Left: $V_2=0$. Right: $V_2=t$.}
  \label{fig:sf-maps}
\end{figure}
\begin{figure}[t]
  \centering
   \includegraphics[width=\columnwidth]{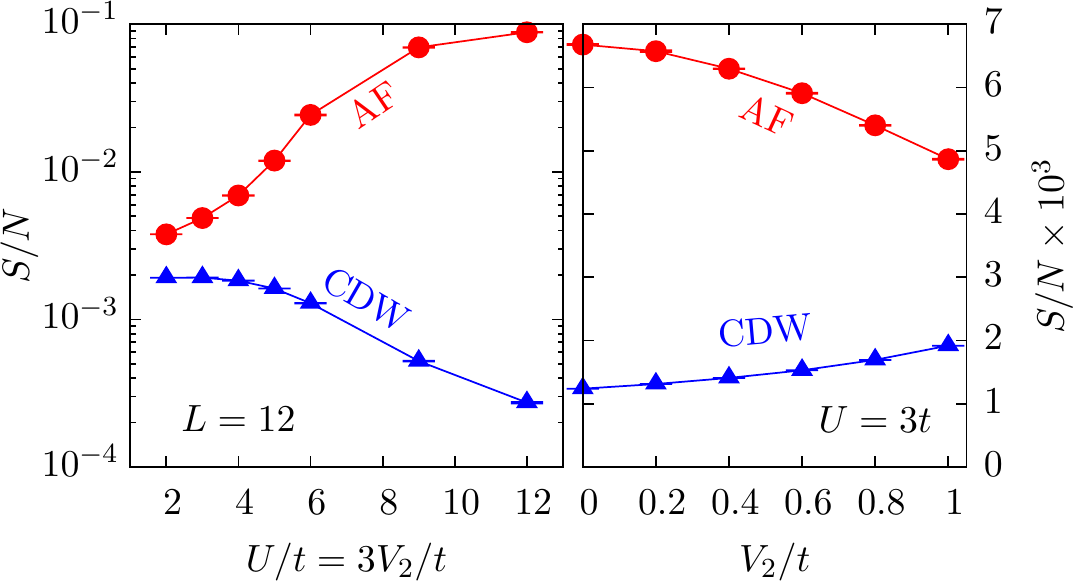}
  \caption{(Color online) Left: Behavior of $S_{\rm AF}$ (red circles) and $S_{\rm CDW}(\bm K)$ (blue triangles) as a function of $U$ at largest possible $V_2=U/3$ for system size $L=12$. Right: Comparison of spin and charge structure factors in the semimetal phase at $U=3t$ and $L=12$.}
  \label{fig:sf-comparison}
\end{figure}
Charge order can be detected by monitoring the corresponding structure factors
\begin{align}
 S^\pm_{\rm CDW}(\bm q)=\frac1N\sum_{\substack{\bm x,\bm y\\s=A,B}} e^{i\bm q(\bm x-\bm y)}\big[\llangle n_{\bm xs}n_{\bm ys}\rrangle\pm \llangle n_{\bm xs}n_{\bm y\bar s}\rrangle\big],
\end{align}
where $\bm x$, $\bm y$ run over all unit cell positions, $s\in\{A,B\}$ is the sublattice index and $\bar s$ denotes the opposite sublattice to $s$. The cumulant is defined as $\llangle OP\rrangle=\langle OP\rangle - \langle O\rangle\langle P\rangle$, for arbitrary operators $O, P$.
As it turns out, the effect of $V_2$-interactions can be most clearly seen in the diagonal structure factor component $S_{\rm CDW}(\bm q)=\frac12[S_{\rm CDW}^+(\bm q) + S_{\rm CDW}^-(\bm q)]$, in which intersublattice correlations are not considered.
In Fig.~\ref{fig:sf-maps}, we show the diagonal charge structure factor component for a $L=18$ system at $U=3t$, comparing the $V_2$-free case with the system at the largest accessible interaction $V_2=t$.
While the former appears devoid of any features, at finite values of $V_2$ a honeycomb-like structure emerges with maxima at wave vector $\bm K$, signaling a tendency towards a charge density wave with such an ordering vector. 
We thus identify $S_{\rm CDW}(\bm K)/N$ as an appropriate order parameter and in Fig.~\ref{fig:sf-comparison} we study its behavior inside the explorable parameter regime alongside the antiferromagnetic structure factor,
\begin{align}
   S_{\rm AF}=\frac1N\sum_{\substack{\bm x,\bm y\\s=A,B}} \big[\langle \bm{S}_{\bm xs}\cdot\bm{S}_{\bm ys}\rangle- \langle \bm S_{\bm xs}\cdot\bm S_{\bm y\bar s}\rangle\big].
\end{align}
Being most promising for the existence of a CDW phase, we immediately focus on the boundary of the sign-problem-free region and perform a scan along the $V_2=U/3$ line.
 One observes that around $U\approx5t$ the AF structure factor begins to grow sizably, while the CDW decays correspondingly (note the logarithmic scale), indicating the onset of the AF phase.
In the semimetal region ($U\lesssim4t$), where $S_{\rm CDW}(\bm K)$ is comparatively large, we can study the influence of $V_2$ at fixed $U=3t$ and find that increasing $V_2$ noticeably weakens AF  and strengthens CDW ordering tendencies. 
However, the AF tendencies remain dominant throughout and a potential phase transition  requires larger values of $V_2$, which we cannot access here.
The finite-size analysis at $U=3t$, $V_2=t$ in Fig.~\ref{fig:honey-sf-fss} confirms this picture and reveals the algebraic behavior of all considered order parameters, and thus indicates that the  semimetallic phase is still present here.

Identifying a possible topological Mott insulator phase proves more difficult than a CDW.
 There exist several approaches to detect the associated topological phase transition in QMC simulations, involving the measurement of Chern numbers\cite{lang2013}, entanglement spectra\cite{grover2013,assaad2015} or strange correlators\cite{wu2015}.
However, from our previous analysis we do not expect such a phase transition to occur in the accessible region, and thus these methods will be of little help here. 

The QSH phase on the honeycomb lattice has been realized by Kane and Mele\cite{kane2005} by adding a NNN  term to the tight-binding description arising from spin-orbit interaction,
\begin{equation}
  \label{eq:kane-mele}
  H_{\rm KM}=H_t+ i\lambda_{\rm SO}\sum_{\llangle ij\rrangle\sigma}(-1)^\sigma v_{ij}c^\dg_{i\sigma}c^{}_{j\sigma},
\end{equation}
where $\lambda_{\rm SO}$ is a measure for the spin-orbit coupling strength, $(-1)^\sigma$ introduces a minus sign for spin-down electrons, and $v_{ij}=\pm1$, depending on whether the hopping path bends to the right or the left.
This term essentially generates circular spin currents inside each triangle of a given sublattice, and stabilizes   a topological phase. 
If an interaction-induced QSH phase is realized in the extended Hubbard model, its onset should be announced by an increase of such spin currents.
We therefore monitor the spin current correlations $\langle j_s(k,l)j_s(m,n)\rangle$ between different NNN bonds, where the spin current operator,
\begin{equation}
  \label{eq:spin-curr}
  j_s(k,l)=-i\sum_\sigma (-1)^\sigma (c^\dg_{k\sigma} c^{}_{l\sigma}-c^\dg_{l\sigma} c^{}_{k\sigma}),
\end{equation}
is measuring the spin current flowing from site $k$ to $l$.

\begin{figure}[t]
  \centering %
  \includegraphics[width=0.46\columnwidth]{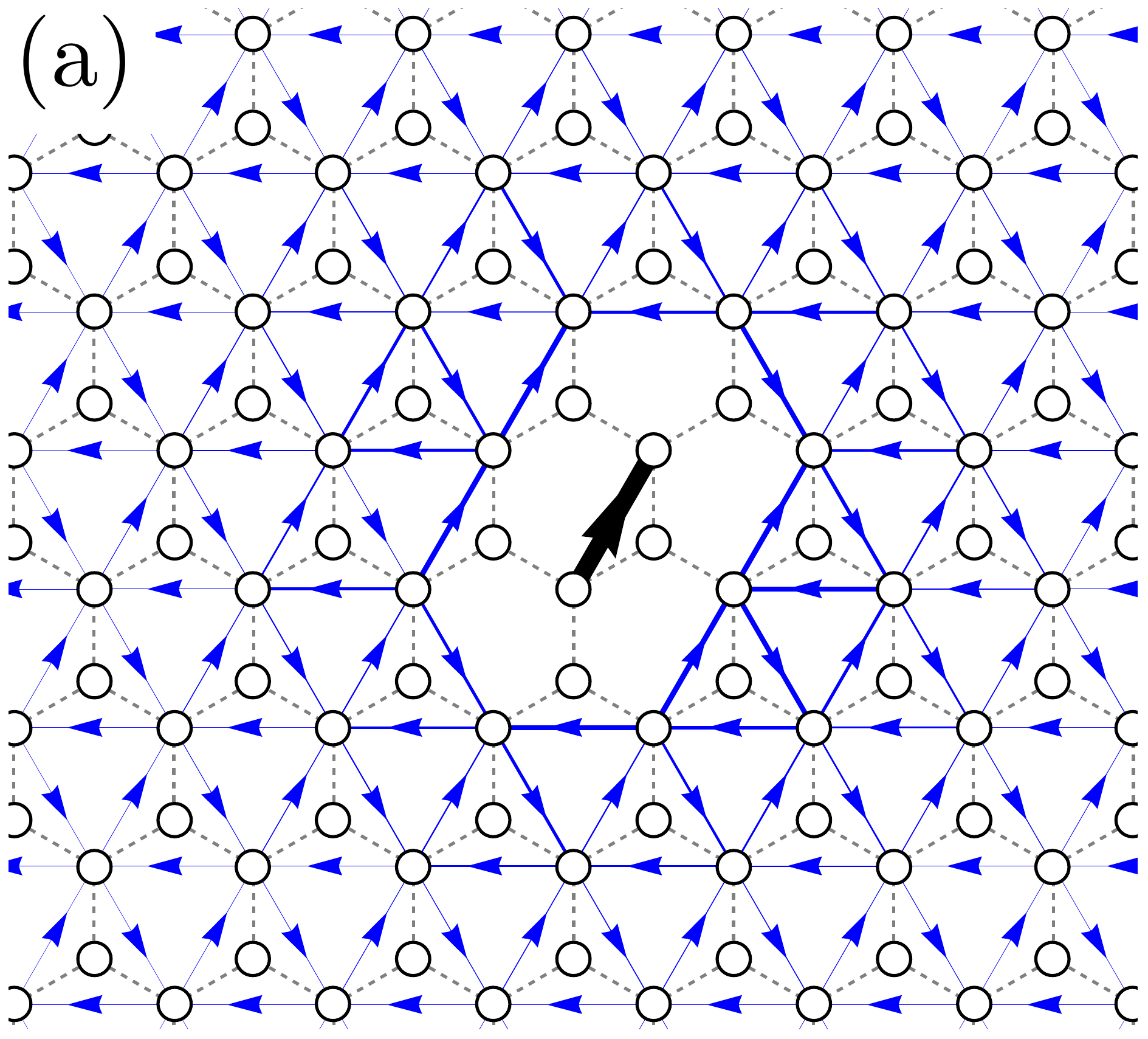}\hfill
  \includegraphics[width=0.51\columnwidth]{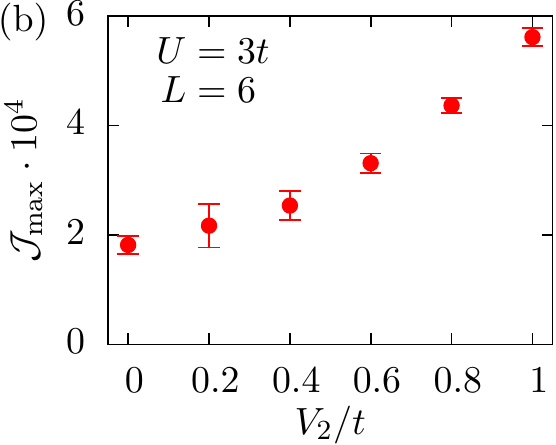}
  \caption{(Color online) Left: Spin current-current correlation pattern within one sublattice for a $L=12$ system at $U=3t$, $V_2=t$. The arrow thickness is proportional to the magnitude and blue arrows denote currents flowing along the arrow direction. The black arrow marks the reference current. Right: Behavior of maximum distance spin current correlations as a function of the extended interaction $V_2$ for a $L=6$ system at $U=3t$.}
  \label{fig:curr-patt}
\end{figure}

In Fig.~\ref{fig:curr-patt}a, the correlations within one sublattice are plotted for a $L=12$ system as close as possible to the potential QSH phase at $U=3t$, $V_2=t$.
As in the Kane-Mele model, the spin currents can be seen to flow circularly inside each triangle, however with the correlation magnitude decaying strongly with distance.
To quantify this decay, we consider the spin current correlation between NNN bonds at the maximum distance on the finite lattice, ${\cal J}_{\rm max}$.
Similar to the structure factors, ${\cal J}_{\rm max}$ shows an algebraic finite-size behavior (Fig.~\ref{fig:honey-sf-fss}), again confirming the perseverance of the semimetallic phase at this point.
In Fig.~\ref{fig:curr-patt}b, we studied ${\cal J}_{\rm max}$ as a function of $V_2$ at $U=3t$, observing a tripling of the spin current correlations in the accessible range, which is a more pronounced increase than shown by the CDW structure factor in Fig.~\ref{fig:sf-comparison}.

In conclusion, our results are compatible with an instability towards a CDW as well as an interaction-driven QSH phase at large $V_2$, but a 
definite statement about the eventually realized phase is not possible, as we cannot access the corresponding parameter regime.

\begin{figure*}[t]
  \centering
  \includegraphics[width=0.98\textwidth]{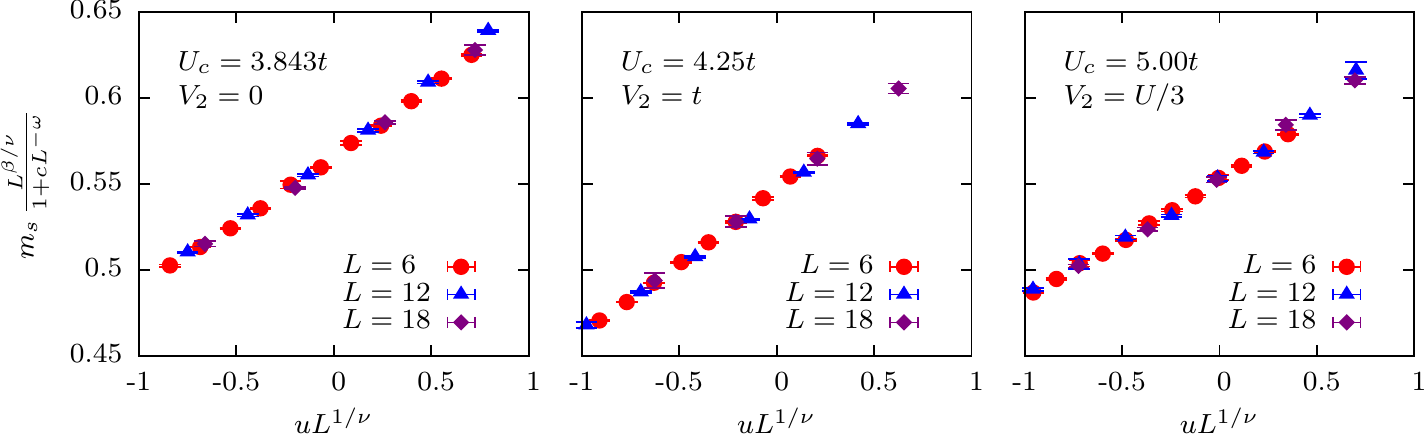}
  \caption{(Color online) Data collapse of the semimetal-to-antiferromagnet transition, assuming a Gross-Neveu-Yukawa critical theory,
  and using the following critical exponents from Otsuka et al.\cite{otsuka2015}: $\nu=1.005$, 
$\beta=0.74$, $\omega=0.55$. The amplitude $c=0.89$ has been determined such that the data in the $V_2=0$ case 
collapses at $U_c=3.843t$, in agreement with the result by Otsuka et al\cite{otsuka2015}. Left: $V_2=0$. Center: $V_2=t$. Right: 
$V_2=U/3$.}
  \label{fig:honey-collapse}
\end{figure*}

\begin{figure}[t]
  \centering
  \includegraphics[width=0.95\columnwidth]{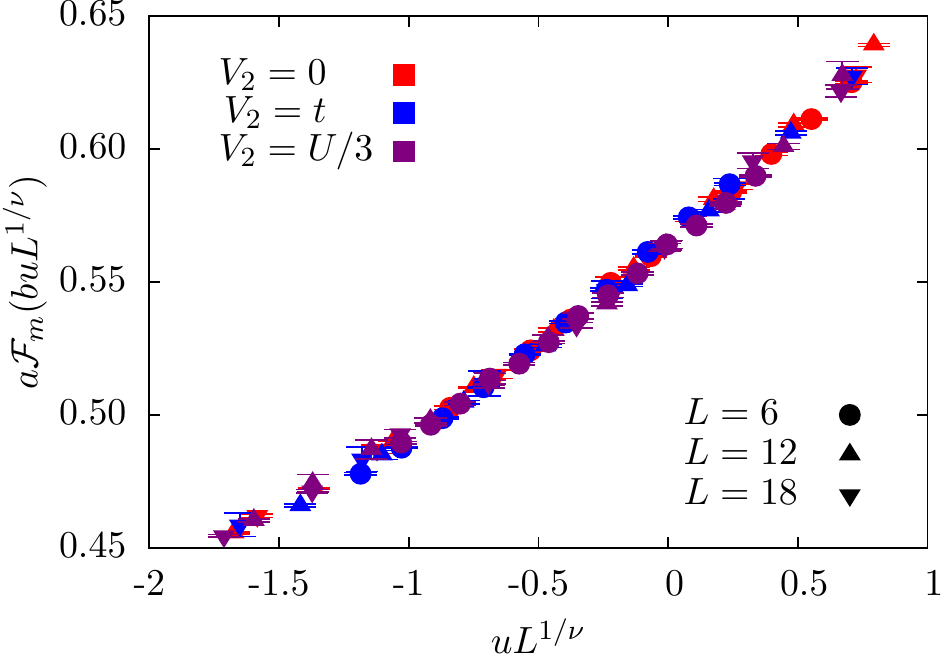}
  \caption{(Color online) Collapse of all three scaling functions from Fig.~\ref{fig:honey-collapse}. The used scale factors are $a=1.13$, $b=1.036$ for $V_2=t$ and $a=0.96$, $b=1.019$ for $V_2=U/3$.}
  \label{fig:honey-scalingfuncs}
\end{figure}

\subsection{Effect of $V_2$ on the metal-insulator transition}
\label{sec:honey-mit}
While no new phases appear within the accessible region, an investigation of the effect of the NNN interaction on the metal-insulator transition may be performed.
For this purpose, we employed a scaling analysis of the staggered magnetization $m_s=\sqrt{S_{\rm AF}/N}$ using the standard finite-size scaling ansatz including leading corrections to scaling,
\begin{equation}
  \label{eq:mag-fss}
  m_s(u,L)=L^{-\beta/\nu}(1+cL^{-\omega})\,\mathcal{F}_m(uL^{1/\nu}),
\end{equation}
with the reduced interaction $u=\frac{U-U_c}{U_c}$ and the scaling function for the magnetization, ${\cal F}_m(u)$.
In the following, we will assume a direct transition from a semimetal to an antiferromagnet, described by the Gross-Neveu-Yukawa theory, with critical exponents $\nu=1.005(5)$, $\beta=0.74(2)$ and the correction exponent $\omega=0.55(4)$, taken from Ref.~\onlinecite{otsuka2015}.

If the rescaled magnetization is plotted against the reduced interaction $u$, the data for different system sizes $L$ will collapse at the correct critical value $U_c$ onto a single curve, which then corresponds to the scaling function ${\cal F}_m(u)$. 
Such an analysis has been perfomed in Fig.~\ref{fig:honey-collapse} for three different values of the extended interaction: At $V_2=0$, $V_2=t$ and also along the edge of the QMC-accessible region, $V_2=U/3$.
In all three cases a data collapse can be achieved, the corresonding critical point $U_c(V_2)$ is shifted to higher 
values with increasing $V_2$, starting from $U_c(V_2=0)=3.843(8)t$ over $U_c(V_2=t)=4.25(2)t$ to 
$U_c(V_2=U/3)=5.00(2)t$.
The resulting phase diagram is shown in Fig.~\ref{fig:honey-phase}. 
Concerning the value of the nonuniversal correction amplitude $c$, we adjusted it such that the data collapse for $V_2=0$ reproduces the critical interaction $U_c=3.843t$ from Ref.~\onlinecite{otsuka2015}, resulting at a value $c=0.89$.  

These results appear consistent with a direct transition described by the Gross-Neveu-Yukawa theory, which is also supported by a closer inspection of the scaling functions for the three transition points:
If they indeed belong to the same universality class, it should be possible to merge their scaling functions by an individual rescaling via $a{\cal F}_m(bu)$, with real numbers $a, b\approx 1$.
This has been done in Fig.~\ref{fig:honey-scalingfuncs}, where the $V_2=0$ scaling functions are kept fixed, while the other scaling functions are rescaled by factors $a=1.13$, $b=1.036$ for $V_2=t$ and by $a=0.96$, $b=1.019$ for $V_2=U/3$, respectively.


\section{Conclusions}
\label{sec:concl}
We presented a method to include nonlocal density interaction in auxiliary field QMC simulations, which allows for sign problem-free calculations inside a restricted parameter region, and  applied it to the half-filled Hubbard model on two different lattice geometries.
In the case of the half-filled Hubbard model on the square lattice bilayer with interlayer repulsion, the ground state phase diagram separates into an ordered region and a disordered regime, depending on the ratio of the interlayer to intralayer hopping.
The nature of the ordered phase is determined by the ratio of interlayer and local repulsions $\Vp/U$, and is 
antiferromagnetic for dominant local interaction and a charge density wave state for strong  interlayer repulsion, 
which we identify within perturbation theory. 
However, the type and  the position of the corresponding phase transition into the charge ordered phase could not 
be determined from the QMC simulations, as it lies outside the accessible parameter regime.
Within the disordered, band insulator phase a peculiar state occurs at equal coupling, $\Vp=U$, for sufficiently large 
interlayer tunneling, which is a direct product of interlayer dimer states of mixed S-Mott and D-Mott character. 
It would be interesting to explore in more detail the transition region between the antiferromagnetic and this 
direct dimer product state, which would however require larger system sizes than those accessible within our investigations. 

For the Hubbard model on the honeycomb lattice with next-nearest-neighbor repulsion $V_2$, only the semimetallic and antiferromagnetic phases appear inside the QMC-accessible region $V_2\le U/3$.
The corresponding phase transition becomes delayed upon increasing $V_2$ and is consistent with
recent estimates of the critical exponents for the chiral-Heisenberg universality class of the Gross-Neveu-Yukawa theory, 
also in the presence of finite next-nearest-neighbor repulsion.
Regarding the nature of the large-$V_2$ phase, results obtained within the semimetallic phase at the largest accessible value of $V_2$ are compatible with a possible instability of the semimetal towards charge order or a topological Mott insulator phase.
A definite statement will require methods that overcome the sign problem also in the large-$V_2$ region.

On a final note, we would like to mention other recent developments~\cite{huffman2014,li2015}, which have resolved the sign problem for simulations of fermionic models with specific extended interactions also in case of an odd number of fermion flavors, such as spinless fermion models with repulsive (attractive) interactions between sites belonging to opposite (equal) sublattices.
Based on the concept of the split orthogonal group, these different approaches have been unified, providing insightful guiding principles for identifying sign problem-free Hamiltonians\cite{wang2015b}, which illustrates the dormant potential of QMC methods.

%
\acknowledgments

We acknowledge discussions with Fakher F.~Assaad, Sylvain Capponi, Akira Furusaki, Martin Hohenadler, Michael M. Scherer,
Sandro Sorella, and Maksim Ulybyshev. We acknowledge support from the Deutsche Forschungsgemeinschaft within grant FOR 1807 and grant WE 
3949/3-1. Furthermore, we acknowledge the allocation of CPU time through JARA-HPC at JSC J\"ulich and at RWTH Aachen.  S.W. thanks the KITP Santa Barbara for hospitality during the program ``Entanglement in Strongly-Correlated Quantum Matter``.
This research was supported in part by the National Science Foundation under Grant No. NSF PHY11-25915.

\bibliography{refs}

\end{document}